\documentclass[a4paper,11pt]{article}
\usepackage{pos}
\usepackage{hyperref}
\usepackage{lineno}

\title{Recent CMS results on exotic resonances}
\manuallySeparateAuthors
\author[a,b]{Jingqing Zhang}
\author[]{ and } %%keep this and the space around and
\author*[a,b]{Kai Yi}
\author{on behalf of the CMS Collaboration}

\affiliation[a]{School of Physics and Technology, Nanjing Normal University,\\
Wenyuan Road No. 1, Nanjing, China}

\affiliation[b]{Department of Physics, Tsinghua University,\\
  Haidian District, Beijing, China}

\emailAdd{jingqing.zhang@njnu.edu.cn}
\emailAdd{yik@fnal.gov}

\abstract{
  Many exotic resonances have been recently observed at the LHC and other 
  experiments. In this report, CMS studies of exotic multiquark states are 
  reported using the data collected in pp collisions at $\sqrt{s}$ = 13 TeV.
}

\FullConference{%
  41st International Conference on High Energy physics - ICHEP2022\\
  6-13 July, 2022\\
  Bologna, Italy
}

\begin{document}
\maketitle

%\linenumbers

\section{Selected CMS contributions to heavy exotic states}
Quantum chromodynamics (QCD) is an important part of the standard model (SM) in particle physics and 
has gotten much support from experimental results as has the rest of the SM. In the QCD framework,
hadrons beyond the quark configurations of $q\bar{q}$ and $qqq$ ($\bar{q}\bar{q}\bar{q}$) are 
`exotic' and are allowed in the QCD theory. Experimental studies on the exotic hadrons will 
help deepen our understanding of QCD.

The CMS experiment~\cite{CMS:2008xjf} at the LHC has performed many important studies
in  hadron spectroscopy and 
%% bph11-006, search for new bottomimum in y1spi+pi-
%% 16-002, search for x(5568) in bs pi+
%% 18-005, b -> jpsi lambda p
%% 19-002, lb -> jpsi lambda phi
%% 18002, Y(1s)mumu
the exotic hadron sector~\cite{CMS:2013fpt,CMS:2013ygz,CMS:2013jru,
CMS:2017hfy,CMS:2019kbn,CMS:2019mny,CMS:2020qwa,CMS:2020eiw}, which include 
the measurement of the $X(3872)$ 
production cross section~\cite{CMS:2013fpt}, 
the confirmation of the $Y(4140)$ 
in $B^{\pm}\rightarrow J/\psi\phi K^{\pm}$ decays~\cite{CMS:2013jru},
and observation of the $B^{0}_{s}\rightarrow X(3872)\phi$ decay~\cite{CMS:2020eiw}
---all
in proton-proton collisions.

Here we present recent results on exotic resonances from the CMS experiment: evidence 
for the $X(3872)$ in heavy-ion collisions~\cite{CMS:2021znk};
observation of the $B^{0}_{s}\rightarrow\psi(2S)K^{0}_{S}$ 
and the $B^{0}\rightarrow\psi(2S)K^{0}_{S}\pi^{+}\pi^{-}$ decays~\cite{CMS:2022cot}, and observation of new 
structures in the $J/\psi J/\psi$ mass spectrum~\cite{CMS-PAS-BPH-21-003}.

\section{Evidence for $X(3872)$ in PbPb collisions}
The $X(3872)$ was first observed by the Belle Collaboration~\cite{Belle:2003nnu}. 
Although its quantum numbers have been determined to be $J^{PC} = 1^{++}$ by the LHCb collaboration~\cite{LHCb:2013kgk},
its nature is still not fully understood. The production and survival of the $X(3872)$ in 
relativistic heavy ion collisions is expected to depend on the $X(3872)$'s internal structure~\cite{Zhang:2020dwn, Wu:2020zbx}.
Therefore, study of the $X(3872)$ production in relativistic heavy-ion collisions provides 
new opportunities to probe the nature of the $X(3872)$.

\begin{figure}[!htbp]
  \centering
  \includegraphics[width=0.6\textwidth]{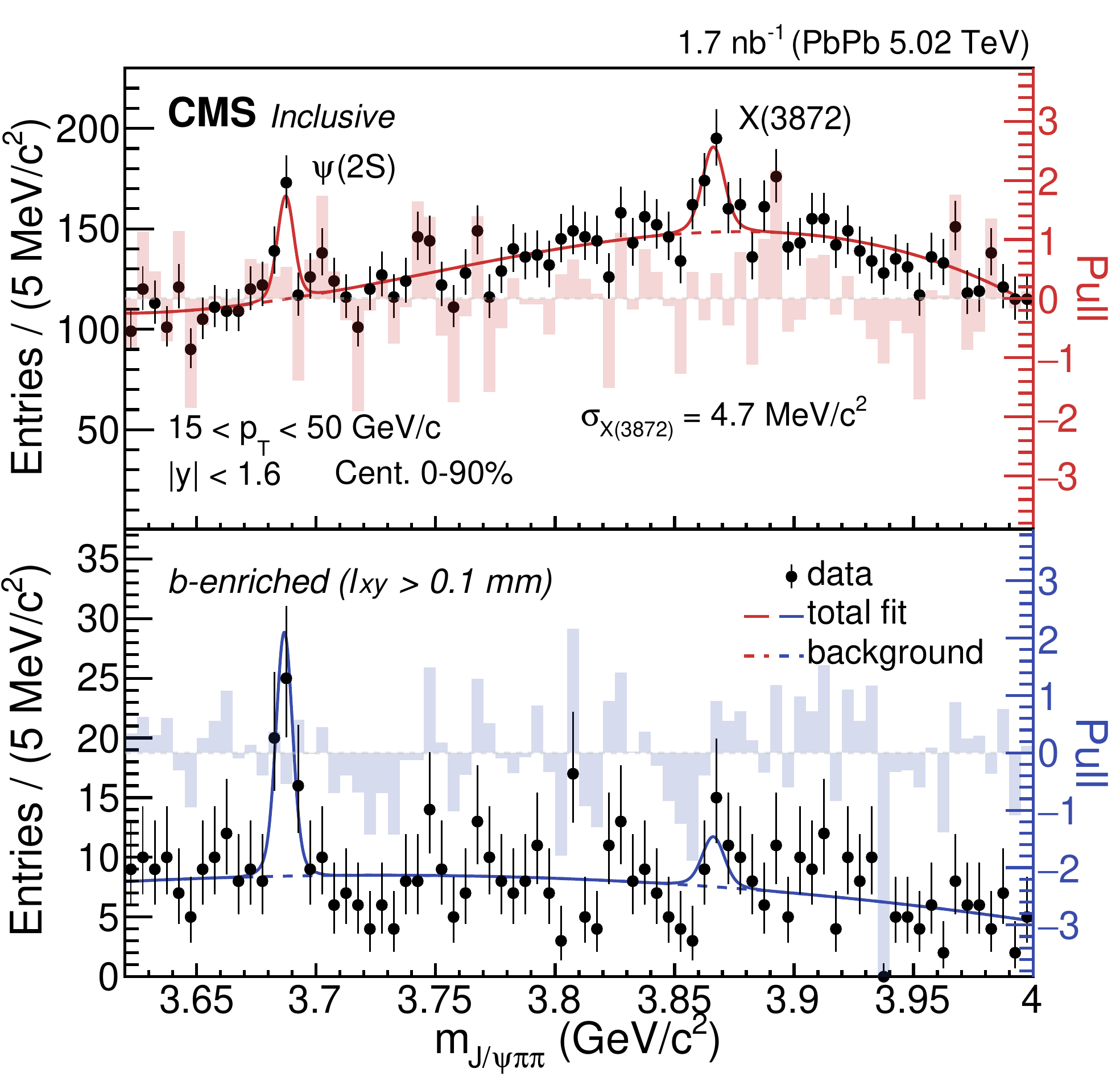}
  \caption{Invariant mass distribution of $m_{\mu\mu\pi\pi}$ 
  %%of the selected candidates in Pb-Pb collisions,
  in Pb-Pb collisions,
  for the inclusive (upper) and $b$-enriched (bottom) samples~\cite{CMS:2021znk}. 
  The vertical lines on points represent statistical 
  uncertainties in the data. The results of the unbinned maximum-likelihood
  fits for the signal + background, and background alone, are also shown by the solid and dashed lines, 
  respectively. The pull distribution is represented by the shaded bars. The $X(3872)$ peak mass resolution,
  $\sigma_{X(3872)}$, calculated at the half-maximum of the signal peak, is also listed
  for reference.}
  \label{fig:x3872pbpb}
\end{figure}

The CMS Collaboration performed a study of $X(3872)$ production in Pb-Pb collisions at $\sqrt{S_{NN}} = 5.02$ TeV 
using 1.7 nb$^{-1}$  sample collected in 2018~\cite{CMS:2021znk}.
The candidates for the $X(3872)$ and $\psi(2S)$ are reconstructed via their decays 
into $J/\psi\pi^{+}\pi^{-}$, where the $J/\psi$ decays into $\mu^{+}\mu^{-}$. 
Figure~\ref{fig:x3872pbpb} shows 
the observed $m_{\mu\mu\pi\pi}$ distribution for the $X(3872)$ and $\psi(2S)$ candidates, where 
the upper plot shows the inclusive sample and the bottom one shows the b-enriched 
(nonprompt dominated, {i.e.} transverse decay length $l_{xy}>0.1$ mm) sample.
The significance of the inclusive $X(3872)$ signal is 4.2 standard deviations. 
%The ratio 
%between the prompt $X(3872)$ and $\psi(2S)$ yields in Pb-Pb collisions, times their branching fractions into 
%$J/\psi\pi^{+}\pi^{-}$, 
The prompt $X(3872)$ to $\psi(2S)$ yield ratio 
is found to be $1.08\pm 0.49(stat) \pm 0.52(syst)$, whereas
the typical value is around 0.1 in pp collisions.

\section{Observation of the $B^{0}_{s}\rightarrow\psi(2S)K^{0}_{S}$ 
and the $B^{0}\rightarrow\psi(2S)K^{0}_{S}\pi^{+}\pi^{-}$ decays}
Multibody decays of the B mesons are well suited to the search for, and  study  of, exotic resonances. For example,
the discovery of $X(3872)$ was in $B\rightarrow K J/\psi\pi\pi$ decays~\cite{Belle:2003nnu}, and that of 
the first charged tetraquark candidate, $Z(4430)^{+}$, was in $B\rightarrow\psi(2S)K\pi^{\pm}$~\cite{Belle:2007hrb}.

\begin{figure}[!htbp]
  \centering
  \includegraphics[width=0.49\textwidth]{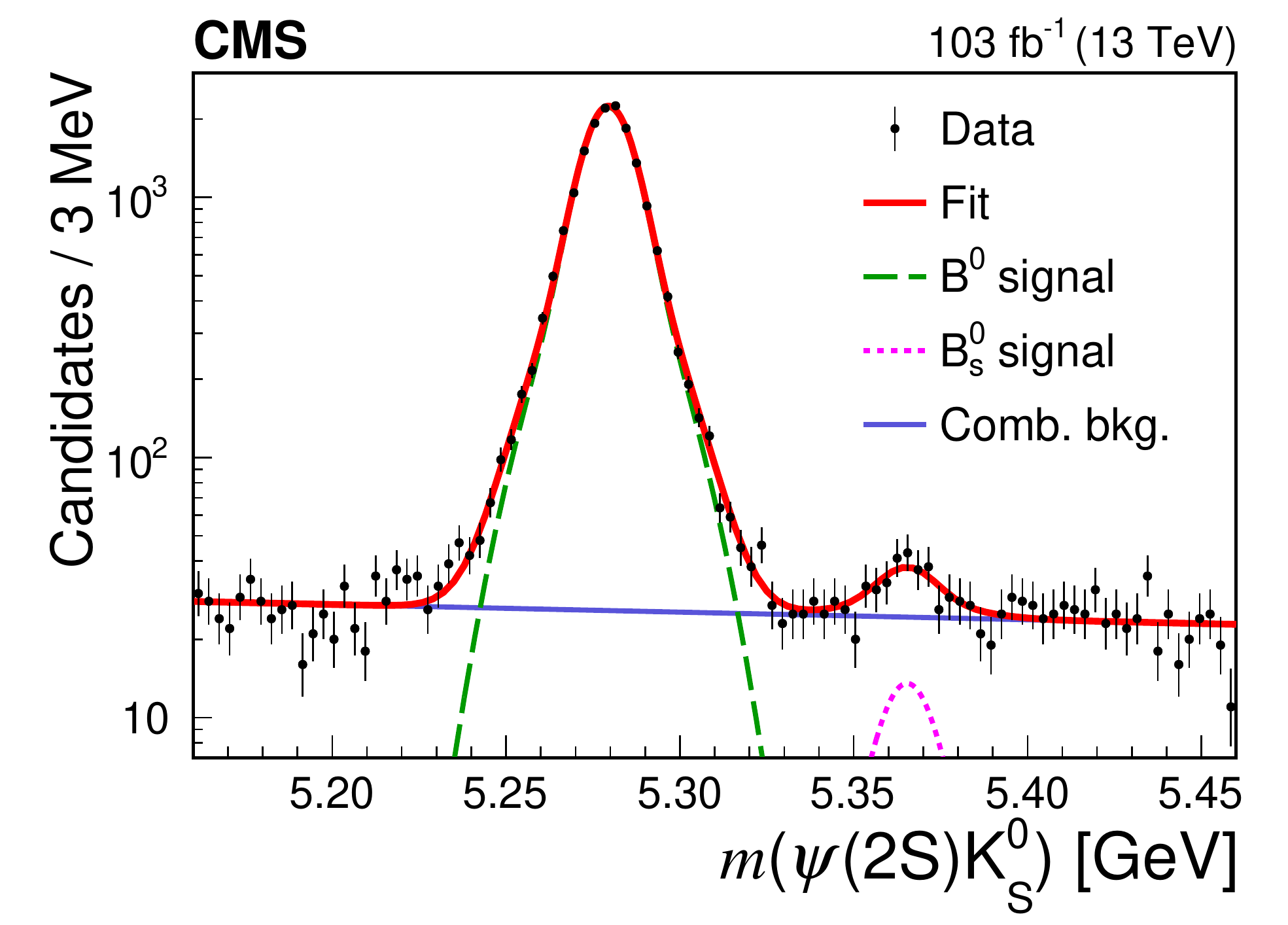}
  \includegraphics[width=0.49\textwidth]{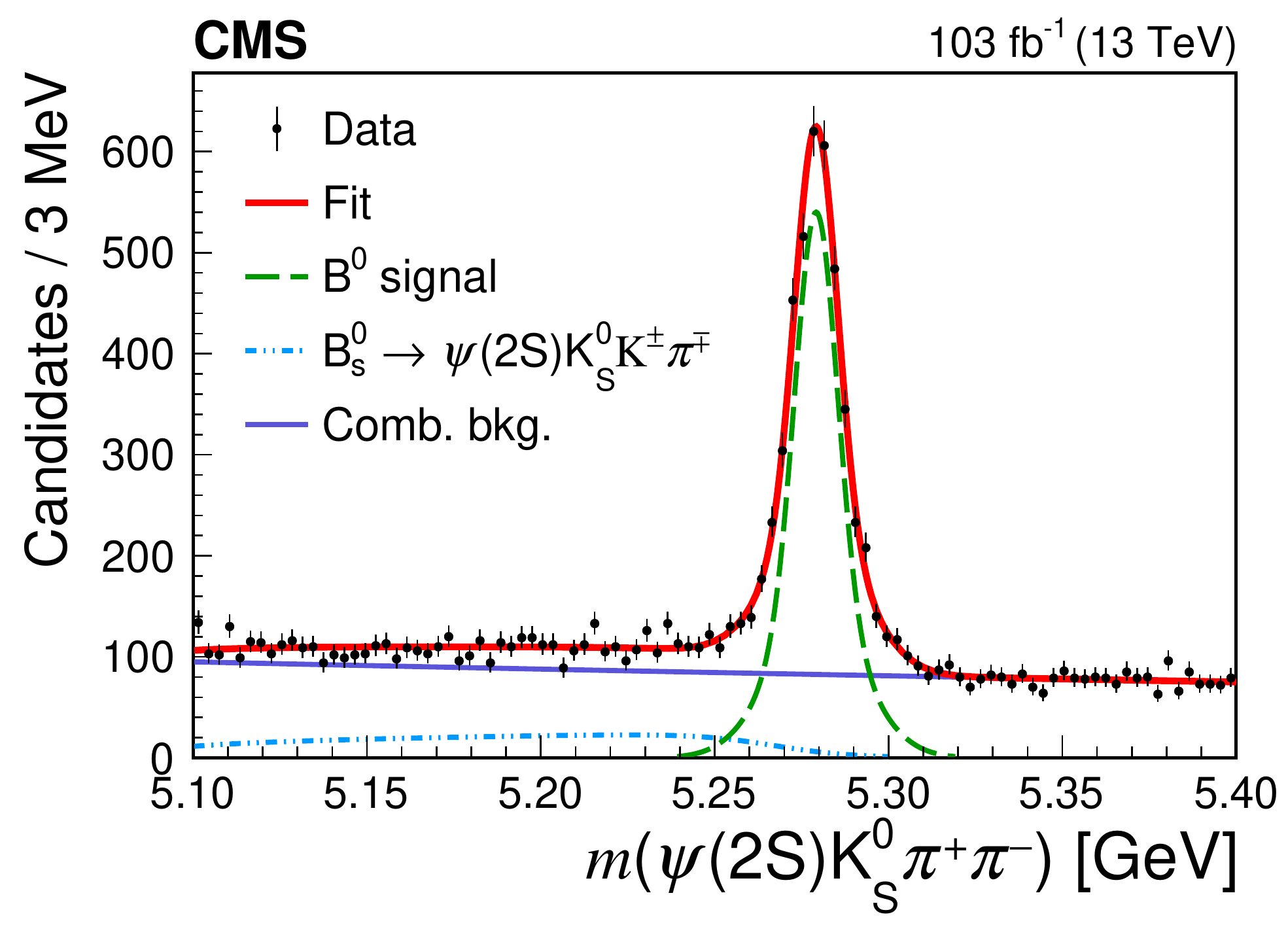}
  \caption{Measured mass distribution of $\psi(2S)K^{0}_{S}$ (left) and $\psi(2S)K^{0}_{S}\pi^{+}\pi^{-}$
  (right) candidates~\cite{CMS:2022cot}.}
  \label{fig:bsb0decay}
\end{figure}

The CMS experiment performed the first measurement of the $B^{0}_{s}\rightarrow \psi(2S)K^{0}_{S}$ 
and $B^{0}\rightarrow\psi(2S)K^{0}_{S}\pi^{+}\pi^{-}$ decays, using a data sample
of proton-proton  collisions at $\sqrt{s} = 13$ TeV,
and an integrated luminosity of 103 fb$^{-1}$,
collected  in 2017 and 2018 ~\cite{CMS:2022cot}.
The $\psi(2S)$ and $K^{0}_{S}$ mesons are reconstructed using their decays into $\mu^{+}\mu^{-}$ 
and $\pi^{+}\pi^{-}$, respectively. 
%The $B^{0}_{s}$ signals are reconstructed using its decay 
%into $\psi(2S)K^{0}_{S}$, while for the $B^{0}\rightarrow\psi(2S)K^{0}_{S}\pi^{+}\pi^{-}$ signals, 
%two additional, oppositely charged, high-purity tracks, assumed to be pions, are included in the 
%signal reconstruction. 
The observed invariant mass distribution of $\psi(2S)K^{0}_{S}$ (left) and 
$\psi(2S)K^{0}_{S}\pi^{+}\pi^{-}$ (right) are shown in Fig.~\ref{fig:bsb0decay}. 
Using the $B^{0}\rightarrow\psi(2S)K^{0}_{S}$ as a reference channel,
the relative branching fractions of 
$B^{0}_{s}\rightarrow\psi(2S)K^{0}_{S}$ and $B^{0}\rightarrow\psi(2S)K^{0}_{S}\pi^{+}\pi^{-}$ decays
are measured to be 
$\mathcal{B}(B^{0}_{s}\rightarrow\psi(2S)K^{0}_{S})/\mathcal{B}(B^{0}\rightarrow\psi(2S)K^{0}_{S})
=(3.33\pm 0.69(stat) \pm 0.11(syst)\pm 0.34(f_{s}/f_{d}))\times 10^{-2}$,
and $\mathcal{B}(B^{0}\rightarrow\psi(2S)K^{0}_{S}\pi^{+}\pi^{-})/
\mathcal{B}(B^{0}\rightarrow\psi(2S)K^{0}_{S}) = 0.480\pm 0.013 (stat) \pm 0.032 (syst)$,
where the last uncertainty in the first ratio corresponds to the uncertainty in the ratio of the production 
cross sections of $B^{0}_{s}$ and $B^{0}$ mesons.
With the currently available, statistics-limited data, 
the 2- and 3- body invariant mass distributions of the 
$B^{0}\rightarrow\psi(2S)K^{0}_{S}\pi^{+}\pi^{-}$ 
decay products do not show significant exotic narrow
structures in addition to the known light meson resonances.

\section{Observation of new structures in the $J/\psi J/\psi$ mass spectrum in pp collisions}
The $X(3872)$ and many other exotic candidates contain two heavy quarks ($c\bar{c}$). 
An analogue to heavy quarkonia would be fully heavy tetraquarks, which have been explored in theoretical models and 
are expected to be experimentally observable. The recent observation of the $X(6900)$ 
decaying into $J/\psi J/\psi$ has been reported by the LHCb Collaboration~\cite{LHCb:2020bwg}.
%, and makes similar studies from other experiments more attractive.
%In this conference,
%the ATLAS and CMS Collaborations released their preliminary results of 
%the $J/\psi J/\psi$ mass spectrum study simultaneously~\cite{ATLAS:2022hhx,CMS-PAS-BPH-21-003}. 
%In this section,
%we present the CMS result briefly~\cite{CMS-PAS-BPH-21-003}.

\begin{figure}[!htbp]
  \centering
  \includegraphics[width=0.49\textwidth]{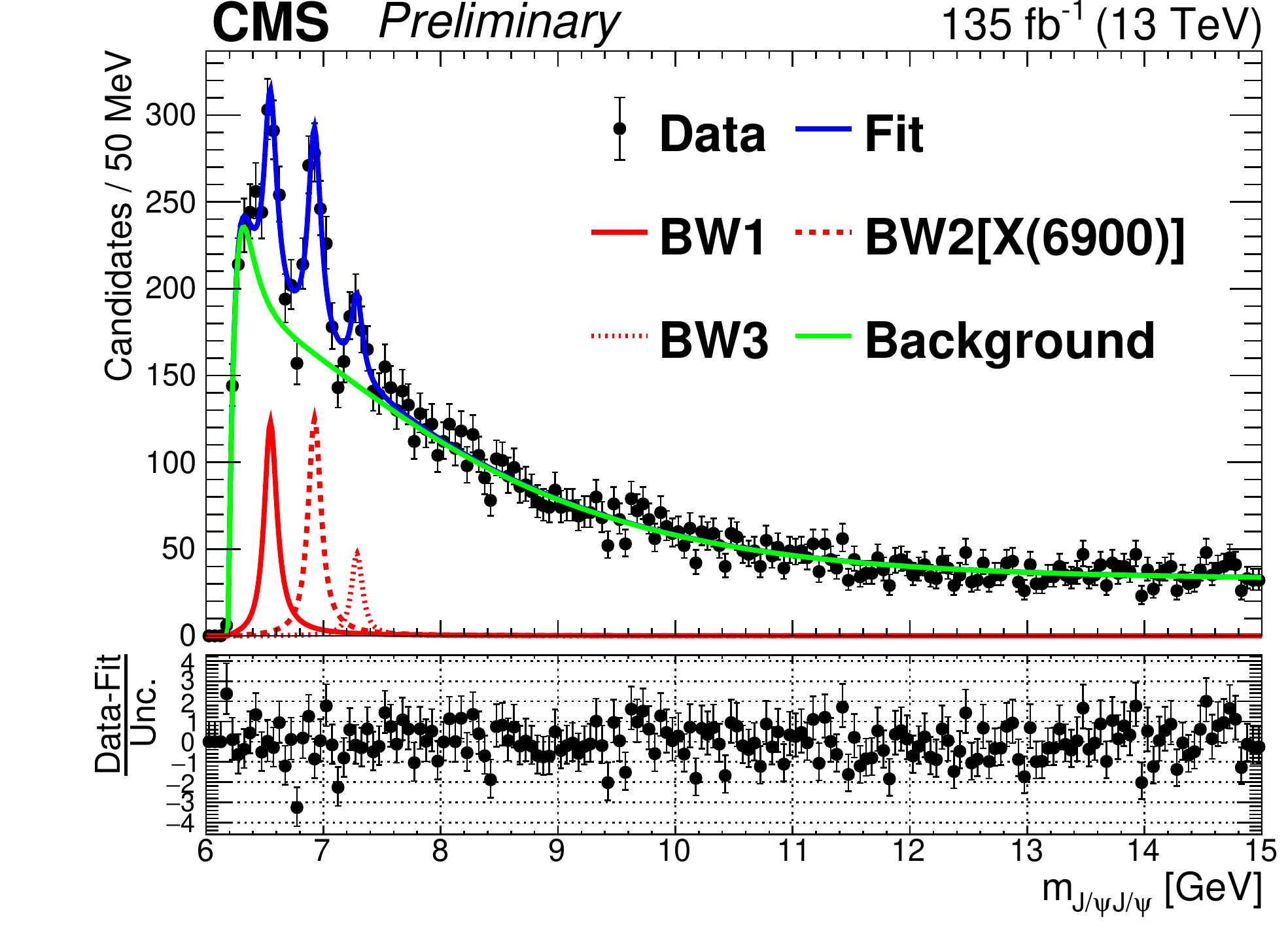}
  \includegraphics[width=0.49\textwidth]{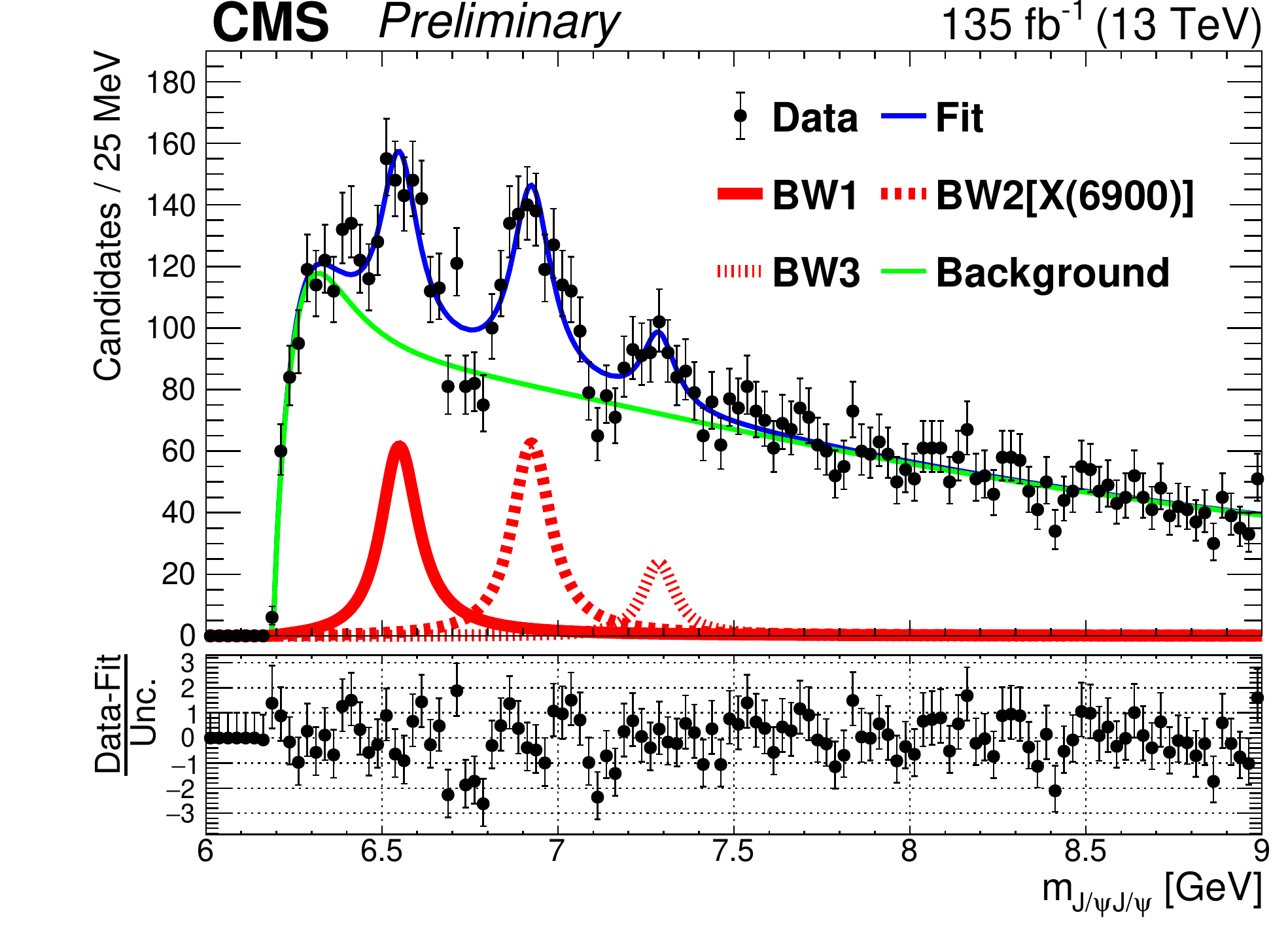}
  \caption{The CMS $J/\psi J/\psi$ mass spectrum with a fit consisting of three signal BW functions 
  and a background model~\cite{CMS-PAS-BPH-21-003}. The left plot shows the fit over the full mass range, and on the right is the 
  same fit expanded by only displaying masses below 9 GeV.}
  \label{fig:jpsijpsi}
\end{figure}

The CMS experiment performed a study of the low-mass region of the $J/\psi J/\psi$ mass spectrum 
in pp collisions, using a data sample  corresponding to 
an integrated luminosity of 135 fb$^{-1}$ at a center-of-mass energy of 13 TeV~\cite{CMS-PAS-BPH-21-003}.
The two $J/\psi$ candidates are reconstructed using their $\mu^{+}\mu^{-}$ mode, 
and the final $J/\psi J/\psi$ mass distribution is shown in Fig.~\ref{fig:jpsijpsi},
where three signal Breit-Wigner structures and a background component are used to fit the distribution.
The statistical significance of the three structures are $6.5\sigma$, $9.4\sigma$, and $4.1\sigma$ for 
$X(6600)$, $X(6900)$ and $X(7300)$, respectively.
The measured masses and widths of three structures are summarized in Table~\ref{tab:jpsijpsi}.

\begin{table}[!htbp]
  \centering
   \begin{tabular}{cccc}
     \hline
     & BW1 & BW2 & BW3 \\
     \hline
     $m$ & $6552\pm10\pm12$ & $6927\pm9\pm5$ & $7287\pm19\pm5$ \\
     $\Gamma$ & $124\pm29\pm34$ & $122\pm22\pm19$ & $95\pm46\pm20$ \\
     $N$ & $474\pm113$ & $492\pm75$ & $156\pm56$ \\
     \hline
   \end{tabular}
   \caption{Summary of the fit results of the CMS $m(J/\psi J/\psi)$ distribution: the mass 
   $m$ and natural width $\Gamma$, in MeV, and the signal yields $N$ are given for three 
   signal structures~\cite{CMS-PAS-BPH-21-003}. The first uncertainties are statistical and the second systematic.}
   \label{tab:jpsijpsi}
\end{table}

\begin{figure}[!htbp]
  \centering
  \includegraphics[width=0.49\textwidth]{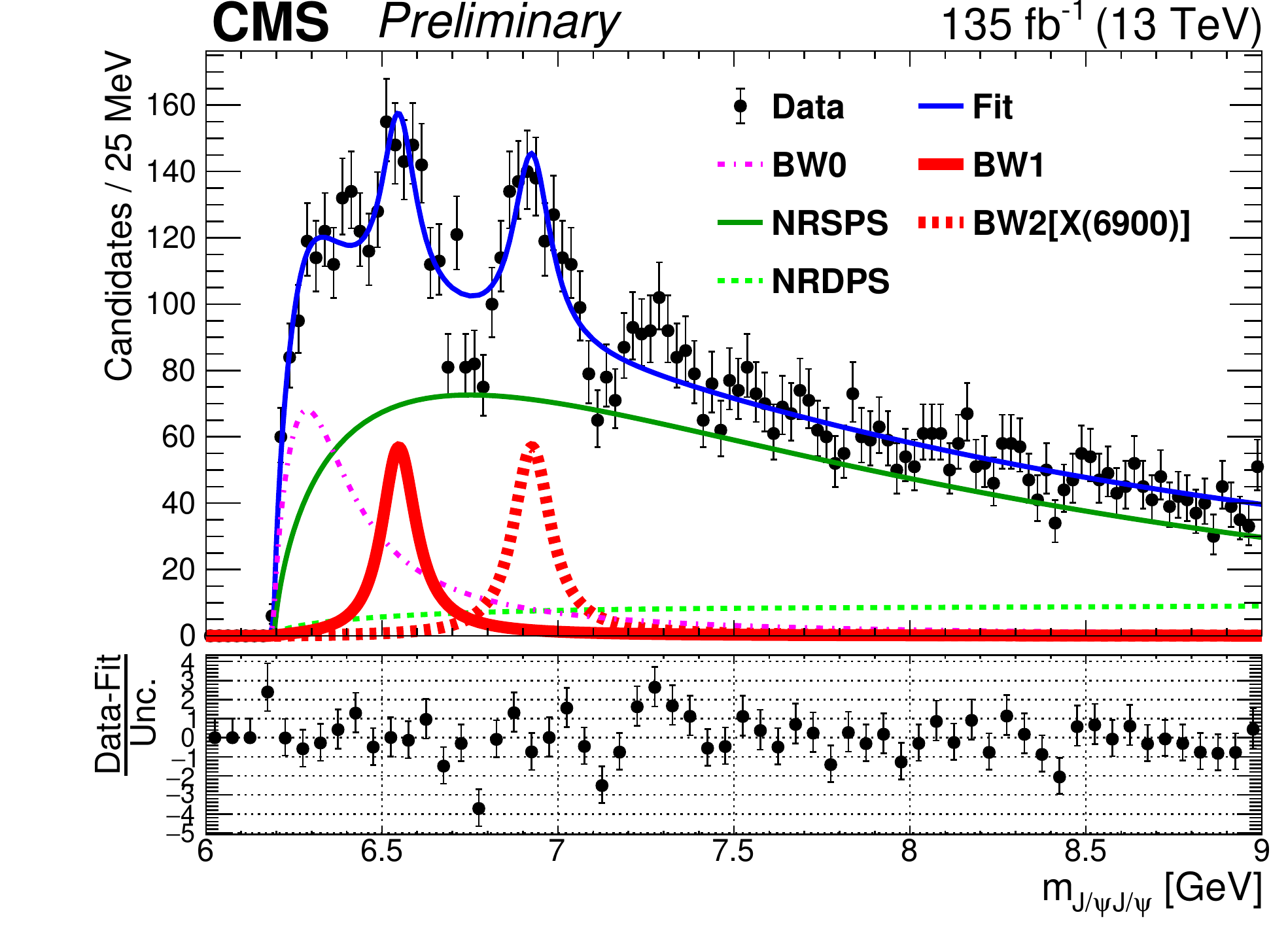}
  \includegraphics[width=0.49\textwidth]{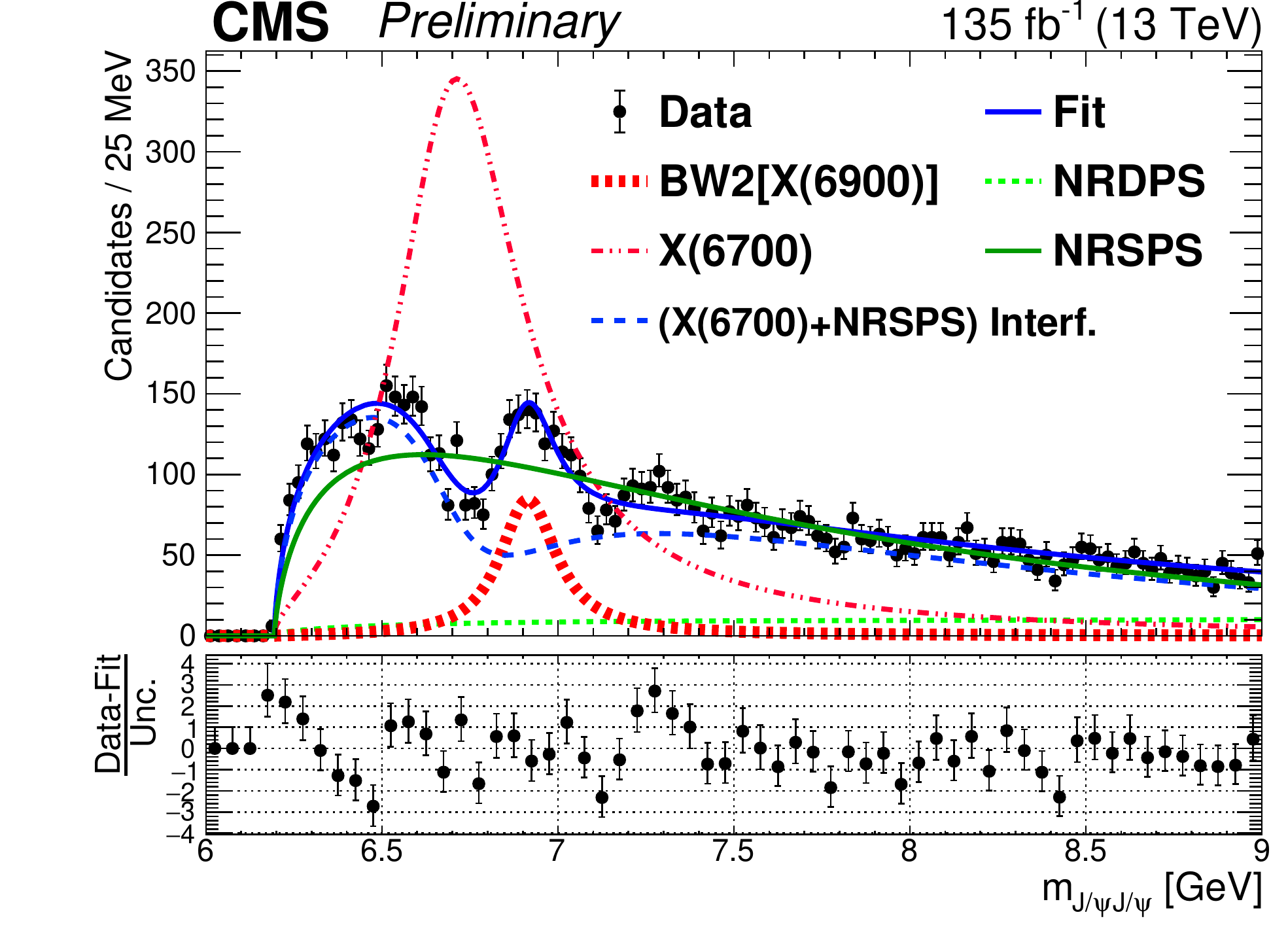}
  \caption{The CMS $J/\psi J/\psi$ mass spectrum~\cite{CMS-PAS-BPH-21-003}. The data are 
  fit using LHCb models: Model I (non-interference) on the left and Model II (interference)
  on the right.}
  \label{fig:lhcbmodel}
\end{figure}

Our $X(6900)$ parameters are in a good agreement with LHCb's non-interference result,
while the $X(6600)$ and $X(7300)$ are new structures.
In order to remove 
potential model dependencies in a comparison of the $X(6900)$ results, we also apply the principal 
two LHCb fit models to the CMS data, but using CMS-specific background shapes.
Figure~\ref{fig:lhcbmodel} shows the application of LHCb's Model I (left, non-interference) 
and Model II (right, non-resonant single parton scattering (NRSPS) interfering with a Breit-Wigner structure -- the $X(6700)$ in our application).
LHCb's Model I consists of the $X(6900)$ signal, NRSPS, non-resonant double parton scattering (NRDPS) and two more BWs -- 
around 6300 (BW0) and 6500 MeV (BW1) -- to account for the threshold enhancement.
LHCb's Model II consists of the $X(6900)$ signal, NRDPS, 
%%and the interference contribution: an interfering Breit-Wigner structure,
%%NRSPS, and their interference. 
and the interference contribution of a Breit-Wigner structure $X(6700)$ and NRSPS.

In both models, the $X(6900)$ parameters are in a good agreement with LHCb's measurements, while our $X(6700)$ in Model II has 
a much larger amplitude and width compared to the LHCb's interfering Breit-Wigner, and none of the LHCb models provide a satisfactory description of our data.

\section{Summary}
In summary, recent CMS results are presented, including 
the first evidence for the $X(3872)$ production in heavy ion collisions, 
the first observation of the
$B^{0}\rightarrow\psi(2S)K^{0}_{S}\pi^{+}\pi^{-}$ and $B^{0}_{s}\rightarrow\psi(2S)K^{0}_{S}$,
and the observation of new structures in $J/\psi J/\psi$ mass spectrum in pp collisions.

%%%%\acknowledgments

\bibliographystyle{JHEP}
\bibliography{myrefs}
%%%\begin{thebibliography}{99}
%%%\bibitem{...}
%%%....
%%%
%%%\end{thebibliography}

\end{document}